\documentstyle[11pt,psfig]{article}
\newcommand{\s}{\mathrm}

\newcommand{\ra}{\rightarrow}
\newcommand{\mn}{\mu \nu}
\newcommand{\be}{\begin{equation}}
\newcommand{\ee}{\end{equation}}

\newcommand{\ba}{\begin{eqnarray}}
\newcommand{\ea}{\end{eqnarray}}

\begin{document}


\begin{center}
{\Large{Photons from Pb + Pb and S + Au collisions\\
at CERN SPS energies}\\
}
\vskip .2in
Sourav Sarkar, Pradip Roy and Jan-e Alam\\

{\it Variable Energy Cyclotron Centre,
     1/AF Bidhan Nagar, Calcutta 700 064
     India}\\

Bikash Sinha\\

{\it Variable Energy Cyclotron Centre,
     1/AF Bidhan Nagar, Calcutta 700 064
     India}\\
{\it Saha Institute of Nuclear Physics,
           1/AF Bidhan Nagar, Calcutta 700 064
           India}\\
\end{center}

\parindent=20pt
\vskip 0.1 in
\begin{abstract}
The effects of the variation of vector meson masses
and decay widths on photon production from hot strongly interacting 
matter formed after Pb + Pb and S + Au collisions 
at CERN SPS energies are considered. It has been shown that the 
present photon spectra measured by WA80 and WA98 Collaborations can not 
distinguish between the formation of quark matter and hadronic matter
in the initial state.

\end{abstract}

\vskip 0.2in

Nucleus-Nucleus collisions at ultra-relativistic energies 
offer a unique opportunity to create and study a new state 
of strongly interacting matter called Quark Gluon Plasma (QGP). 
Photons and dileptons can probe the entire volume of the plasma
without almost any interaction and as such are better markers
of space time history of the evolving matter~\cite{pr}.
However, apart from QGP, photons can originate 
from the primary interactions among the partons of the colliding nuclei,
which dominate the high momentum region of the spectra, and these photons 
could be evaluated reliably by applying perturbative QCD (pQCD).
At smaller values of the 
transverse momentum, meson decays (mainly $\pi^{0}$ and also 
$\eta$) dominate 
the spectrum. Due to their long life time $\pi^0$ decays 
into two photons outside the hot zone and 
photons originating from this decay 
can be reconstructed through invariant mass analysis. 
But there is 
no method by which the thermal photons from hadronic reactions and 
decays within the hot zone of the thermalised system can be identified
experimentally. In an ideal scenario where all the photons 
from $\pi^0$ decays are reconstructed and the photons from hard 
QCD processes are identified and subtracted from the data,
then only thermal photons will be 
left in the data. 

Irrespective of whether QGP is formed or not, hadronic 
matter (HM) formed in Ultra-relativistic Heavy Ion Collisions 
(URHIC) is expected to be in a highly excited
state of very high temperature and/or density. Thus it is
of primary importance to understand the change in hadronic 
properties {\it e.g.} mass, life time etc at finite temperature
and density. One  of the most important aspects, 
spontaneously broken chiral symmetry, a property of hadrons in their 
ground state, is expected to be restored at high temperature, 
which should manifest itself in the thermal shift of hadronic masses
as well as decay widths. Changes in the hadronic properties 
could be probed most efficiently
by studying the thermal spectrum of real and virtual (dilepton pairs)
photons. The thermal photon yield 
from S + Au collisions has been studied by many 
authors~\cite{arbex,dks,dumitru,neumann} 
without taking medium effects into account.
In this work we evaluate the transverse momentum
distribution of photons emitted from a strongly interacting system
with initial conditions expected to be realised at CERN SPS energies
for Pb + Pb and S + Au collisions, taking in-medium effects on hadronic
properties into account.

In a phase transition scenario, thermal photons originate both from 
QGP and hadronic phase as the latter is  realised when the temperature of
the system cools down to the critical temperature ($T_c$) due to
expansion. However, if the system 
does not go through a phase transition, then obviously,
the thermal photons originate from hadronic interactions only.
We have studied the thermal photon spectra from both the  
scenarios.

The thermal emission rate 
of real photons can be expressed in terms of the trace of the retarded 
photon self energy ($\Pi_{\mn}^R$)at finite temperature~\cite{gale}
\be
E\frac{dR}{d^3p}=-\frac{2g^{\mn}}{(2\pi)^3}{\s {Im}}\Pi_{\mn}^R\,(p)\frac{1}
{e^{E/T}-1}
\label{photrate1}
\ee
where $g_{\mn}$ is the metric tensor and $T$ is the temperature of the
thermal medium.
In the quark matter (QM) the lowest order contribution 
to the trace of the imaginary part of the retarded 
self energy ${\s {Im}}\Pi_{\mn}^R\,(p)$ comes from the two loop
diagrams corresponding to QCD Compton and annihilation processes,
the total rate for which is given by~\cite{solfrank}, 
\be
E\frac{dR}{d^3p}=\frac{5}{9}\frac{\alpha\alpha_s}{2\pi^2}\,T^2
\exp(-E/T)\,\ln(\frac{0.2317E}{4\pi\,\alpha_sT})
\label{photrate2}
\ee
where $\alpha$ is the fine structure constant, $\alpha_s=\,6\,\pi/
(33-2n_f)\ln(8\,T/T_c)$~\cite{karsch} 
is the strong coupling constant.

In the hadronic matter (HM) an exhaustive set of hadronic reactions 
and vector meson decays involving $\pi$, $\rho$, $\omega$ and
$\eta$ mesons have been considered. 
It is well known~\cite{kapusta} 
that the reactions $\pi\,\rho\,\ra\, \pi\,\gamma$ , 
$\pi\,\pi\,\ra\, \rho\,\gamma$ , $\pi\,\pi\,\ra\, \eta\,\gamma$ , 
$\pi\,\eta\,\ra\, \pi\,\gamma$ , and the decays $\rho\,\ra\,\pi\,\pi\,\gamma$
and $\omega\,\ra\,\pi\,\gamma$ are the most important channels 
for photon production from hadronic matter in the
energy regime of our interest. The rates for these processes could be 
evaluated from the imaginary part of the two loop photon self energy
involving various mesons.
Recently it has been shown~\cite{halasz} that 
the role of intermediary $a_1$ in the photon producing reactions is less
important than thought earlier~\cite{xiong,song}. 
In the present work we have neglected $a_1$ in the intermediate state.
 
The full interaction 
Lagrangian density and the relevant matrix elements for these processes 
have been
given in Refs.~\cite{ss,pkr}, we do not repeat those here. 
The photon emission rate from 
HM can not be expressed in a closed form as in Eq.~(\ref{photrate2}) due
to the complexities arising from the nature of the hadronic
interactions and reaction kinematics.

To study the medium effects on the transverse momentum distribution 
of photons from URHIC we need two more ingredients. Firstly, we require
the variation of masses and decay widths with temperature, because
the invariant matrix element for photon production suffer in-medium
modifications through the temperature dependent masses and
widths of the participants. 
As the hadronic masses and decay widths enter directly in the 
count rates of electromagnetically interacting particles, the finite
temperature and density effects in the cross sections, particularly
in the HM are very important in URHIC. In our earlier 
calculations~\cite{pkr} 
we have studied the finite temperature and density effects on hadronic 
properties by applying finite temperature field theory
within a framework of an effective Lagrangian approach. 
The variation of nucleon, rho and omega masses and the decay width
of rho with temperature could be parametrized as 
\be
m_N^\ast/m_N=1-0.0264(T/T_c)^{8.94}\nonumber
\ee
\be
m_\rho^\ast/m_\rho=1-0.1268(T/T_c)^{5.24}\nonumber
\ee
\be
m_\omega^\ast/m_\omega=1-0.0438(T/T_c)^{7.09}\nonumber
\ee
\be
\Gamma_\rho^\ast/\Gamma_\rho=1+0.6644(T/T_c)^4-0.625(T/T_c)^5
\ee
where asterisk indicates effective mass in the medium and $T_c=0.16$ GeV. 
In the rho width we have included the Bose enhancement
effects~\cite{pkr}. 
Note that in our calculation the nucleon, rho and omega masses decrease
differently; we do not observe any universal scaling law~\cite{mrho}.
Effects of scaling mass variation~\cite{mrho} on photon spectra
has recently been studied by Song et al~\cite{csong}.

The observed photon spectrum originating from an expanding 
QGP or hadronic matter is obtained by convoluting the static
(fixed temperature) rate, as given by Eq.~(\ref{photrate1}), with 
expansion dynamics. 
Therefore, the second ingredient required for our calculations is 
the description
of the system undergoing rapid expansion from its initial formation
stage to the final freeze-out stage.
In this work we use Bjorken-like~\cite{bjorken}
hydrodynamical model for the isentropic expansion of the matter
in ($1 + 1$) dimension.
For the QGP sector we use simple bag model equation of state (EOS) with
two flavour degrees of freedom. The temperature in the QGP phase evolves
according to Bjorken scaling law $T^3\,\tau=T_i^3\tau_i$.

In the hadronic phase we have to be more careful about the presence
of heavier particles and their change in masses due to finite temperature
effects.
The hadronic phase consists of $\pi$, $\rho$, $\omega$, $\eta$ and $a_1$ 
mesons and nucleons. The nucleons and heavier mesons may play an important
role in the EOS in a scenario where mass of the hadrons decreases
with temperature. 

The energy density and pressure
for such a system of mesons and nucleons is given by,
\be
\epsilon_H=\sum_{i=mesons} \frac{g_i}{(2\pi)^3} 
\int d^3p\,E_i\,f_{BE}(E_i,T)
+\frac{g_N}{(2\pi)^3} 
\int d^3p\,E_N\,f_{FD}(E_N,T)
\ee
and
\be
P_H=\sum_{i=mesons} \frac{g_i}{(2\pi)^3} 
\int d^3p\frac{p^2}{3\,E_i}f_{BE}(E_i,T)
+\frac{g_N}{(2\pi)^3} 
\int d^3p\frac{p^2}{3\,E_N}f_{FD}(E_N,T)
\ee
where the sum is over all the mesons under consideration and $N$ stands
for nucleons and $E_i=\sqrt{p^2 + m_i^2}$.          
The entropy density is given by
\be
s_H=\frac{\epsilon_H+P_H}{T}\,\equiv\,4a_{\s{eff}}(T)\,T^3
= 4\frac{\pi^2}{90} g_{\s{eff}}(m^\ast(T),T)T^3
\label{entro}
\ee
where  $g_{\s{eff}}$ is the effective statistical degeneracy.

Thus, we can visualise the finite mass of the hadrons
having an effective degeneracy $g_{\s{eff}}(m^\ast(T),T)$. The variation 
of temperature from its initial value  $T_i$ to final value 
$T_f$ (freeze-out temperature) with proper time ($\tau$) is governed 
by the entropy conservation 
\be
s(T)\tau=s(T_i)\tau_i
\label{entro1}
\ee
The initial temperature of the system is obtained by solving
the following equation self consistently
\be
\frac{dN_\pi}{dy}=\frac{45\zeta(3)}{2\pi^4}\pi\,R_A^2 4a_{\s{eff}}T_i^3\tau_i
\ee
where $dN_\pi/dy$ is the total pion multiplicity, $R_A$ is the radius
of the system, $\tau_i$ is the initial thermalisation time and 
$a_{\s{eff}}=({\pi^2}/{90})\,g_{\s{eff}}(m^\ast(T_i),T_i)$. 
The change in the expansion dynamics
as well as the value of the initial temperature due
to medium effects enters the calculation of the
photon emission rate through the effective statistical degeneracy.

We consider Pb + Pb collisions at CERN SPS energies. 
If we assume that the matter is formed in the QGP phase
with two flavours ($u$ and $d$), then $g_k=37$.
Taking $dN_{\pi}/dy=600$
as measured by the NA49 Collaboration~\cite{npa610} 
for Pb + Pb collisions, we obtain $T_i=180$ MeV for
$\tau_i=1$ fm/c. The system takes a time $\tau_Q=T_i^3\tau_i/T_c^3$ 
to achieve
the critical temperature of phase transition ($T_c$=160 MeV in our case).
In a first order phase transition scenario the system remains in the mixed
phase up to a time $\tau_H\,=\,r\,\tau_Q$, 
{\it i.e.} $T$ remains at $T_c$ for an interval $\tau_H-\tau_Q$,
where $r$ is the ratio of 
the statistical degeneracy in QGP to hadronic phase. At $\tau_H$ the 
system is fully converted to hadronic matter and remains in this 
phase up to a proper time $\tau_f$. We have taken
$T_f=130$ MeV in our calculations.

\begin{figure}
\centerline{\psfig{figure=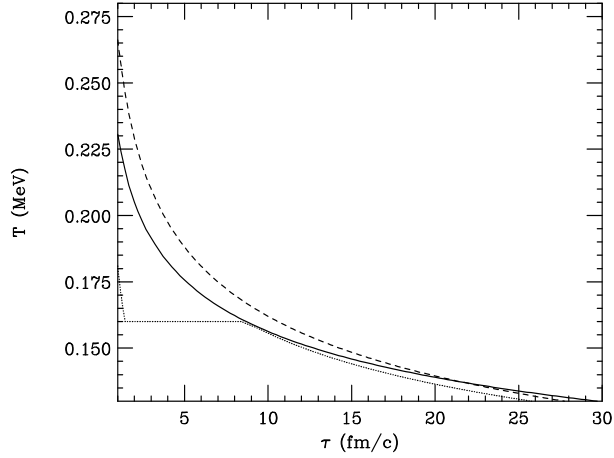,height=6cm,width=8cm}}
\caption{Variation of temperature with proper time. 
The dotted line indicates the cooling law in a first
order phase transition scenario. The solid (dashed) line
represents temperature variation in a `hadronic scenario'
with (without) medium effects on the hadronic masses.
}
\label{cooling}
\end{figure}

In Fig.~(\ref{cooling}) we demonstrate the variation of temperature 
with proper time for different initial conditions. 
The dotted line indicates the scenario where QGP is formed initially 
at $T_i=180$ MeV and cools down according to 
Bjorken law up to a temperature $T_c$ at which a phase transition
takes place; it remains constant at $T_c$ up to a time
$\tau_H=8.4$ fm/c after which the temperature decreases as 
$T=0.241/\tau^{0.19}$ to a temperature $T_f$. 
If the system is considered to be formed in the hadronic phase 
the initial temperature is obtained as $T_i=230$ MeV (270 MeV)
when in-medium effects on the hadronic masses are taken into account
(ignored), the corresponding cooling laws are
$T=0.230/\tau^{0.169}$ ($T=0.266/\tau^{0.2157}$) 
are displayed in Fig.~(\ref{cooling}) by solid and dashed lines respectively. 
The above parametrizations of the cooling law in the hadronic phase
have been obtained by solving Eq.~(\ref{entro1}) self consistently.

Obtaining the finite temperature effects on hadronic properties
and the cooling law we are ready to evaluate the photon spectra
from an $(1+1)$ dimensionally expanding system.
The transverse momentum distribution of photons in a first order
phase transition scenario is given by,

\begin{eqnarray}
E\frac{dN}{d^3p}&=&\pi\,R_A^2\,\int\left[
\left(E\frac{dR}{d^{3}p}\right)_{QGP}
\Theta(\epsilon-\epsilon_{Q})\right.\nonumber\\
& &+\left[\left(E\frac{dR}{d^{3}p}\right)_{QGP}\frac{\epsilon-
\epsilon_{H}}
{\epsilon_{Q}-\epsilon_{H}}\right.\nonumber\\
& &+\left.\left(E\frac{dR}{d^{3}p}\right)_{H}\frac{\epsilon_{Q}-
\epsilon}
{\epsilon_{Q}-\epsilon_{H}}\right]
\Theta(\epsilon_{Q}-\epsilon)
\Theta(\epsilon-\epsilon_{H})\nonumber\\
& &+ 
\left.\left(E\frac{dR}{d^{3}p}\right)_{H}\Theta(\epsilon_{H}-
\epsilon)\right]
\tau\,d\tau\,d\eta 
\end{eqnarray}
where $\Theta(M)=\Theta(\epsilon_Q-\epsilon)\Theta(\epsilon-\epsilon_H)$,
$\epsilon_Q$ ($\epsilon_H$) is the energy density in the 
QGP (hadronic) phase at $T_c$, $\eta$ is the space time rapidity,
$R_A$ is the radius of the nuclei and $\Theta$ functions are 
introduced to get the contribution from individual phases.

\begin{figure}
\centerline{\psfig{figure=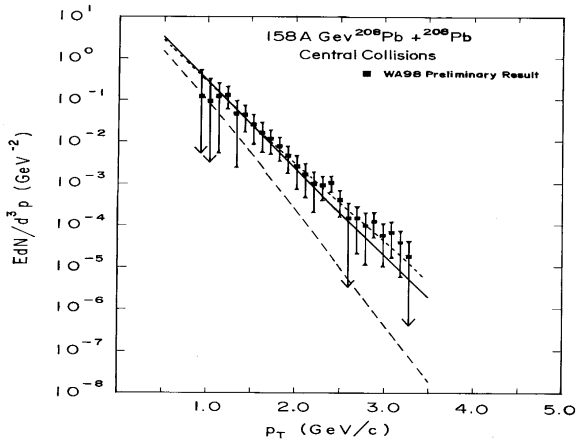,height=6cm,width=8cm}}
\caption{Total thermal photon yield in Pb + Pb central collisions
at 158 GeV per nucleon at CERN SPS. The long-dash line shows the results
when the system is formed in the QGP phase 
with initial temperature $T_i=180$ MeV
at $\tau_i=1$ fm/c. The critical temperature for phase transition is
taken as 160 MeV. The solid
(short-dash) line indicates photon spectra  
when hadronic matter formed in the initial state at $T_i=230$ MeV
($T_i=270$ MeV) at $\tau_i=1$ fm/c 
with (without) medium effects on hadronic masses and decay widths.
}
\label{wa98}
\end{figure}

In Fig.~(\ref{wa98}) we compare our results of transverse momentum
distribution of photons  with the preliminary results of WA98
Collaboration~\cite{wa98con}. 
The experimental data represents the photon spectra
from Pb + Pb collisions at 158 GeV per nucleon at CERN
SPS energies. The transverse momentum distribution of photons originating
from the `hadronic scenario' (matter formed in the hadronic
phase) with (solid line) and without(short-dash line) 
medium modifications of vector mesons outshine the photons 
from the `QGP scenario' (matter formed in the QGP phase, indicated 
by long-dashed line) for the entire range of $p_T$.
Although photons from `hadronic scenario' with medium
effects on vector mesons shine less bright than those from
without medium effects for $p_T>2$ GeV, it is not possible
to distinguish clearly between one or the other on the basis of the 
experimental data.  

\begin{figure}
\centerline{\psfig{figure=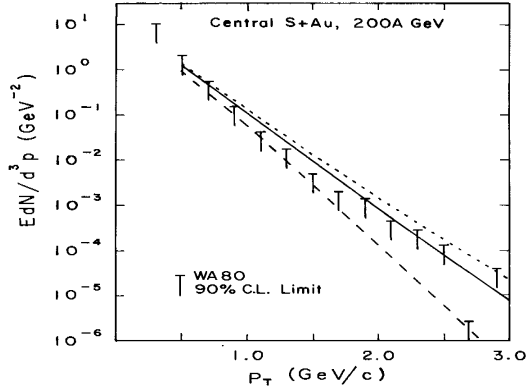,height=6cm,width=8cm}}
\caption{Total thermal photon yield in S + Au central collisions
at 200 GeV per nucleon at CERN SPS. The long-dash line shows the results
when the system is formed in the QGP phase 
with initial temperature $T_i=190$ MeV
at $\tau_i=1.2$ fm/c. The critical temperature for phase transition is
taken as 160 MeV. The solid
(short-dash) line indicates photon spectra  
when hadronic matter formed in the initial state at $T_i=230$ MeV
($T_i=270$ MeV) at $\tau_i=1.2$ fm/c 
with (without) medium effects on hadronic masses and decay widths.
}
\label{wa80}
\end{figure}

In Fig.~(\ref{wa80}) we compare thermal photon spectra with 
the upper bound of WA80 Collaboration~\cite{wa80prl}. The experimental
data stands for S + Au collisions at 200 GeV per nucleon
at SPS. In this case the pion multiplicity, $dN_\pi/dy=225$.
Our calculation shows that maximum number of photons originate
from the `hadronic scenario' 
when the medium effects on vector mesons are ignored.
For such a scenario the photon spectra has just crossed the upper
bound of the WA80 data and more likely such a scenario is not realised
in these collisions. This is in line with the analysis of the
preliminary WA80 data of Ref.~\cite{dks}. It should be noted for 
completeness that the temperature variation of the degeneracy factor
was not considered in Ref.~\cite{dks} leading to 
quantitative differences in the theoretical predictions between 
the present work and Ref.~\cite{dks}. 
Photons from `hadronic scenario'
with medium modifications of vector meson properties outshine
those from the `QGP scenario' for the entire $p_T$ range.
Considering the experimental uncertainty, 
no definitive conclusion can be drawn in favour of any particular
scenario.

We have compared the experimental data on photon spectra from S + Au 
and Pb + Pb collisions at 200 GeV and 158 GeV per nucleon respectively
with different initial conditions. In case of Pb + Pb collisions 
photons from hadronic scenario dominates over the photons from
QGP scenario for the entire $p_T$ domain. 
But in the hadronic scenario the photon spectra evaluated with
and without in-medium properties of vector mesons describe 
these data reasonably well. Hence the 
transverse photon spectra at present do not allow us to decide between
an in-medium dropping mass and a free mass scenario. 
For S + Au collisions the photon spectra  evaluated
with free masses seems to exceed the experimental upper bound.
However, the photon spectra 
obtained by assuming hadronic matter (with in-medium effects or 
otherwise) in the initial state 
outshines the spectra evaluated with first order phase transition.
Considering the experimental uncertainty, it is not possible
to state, which one, between the two is compatible with the data.
Experimental data with better statistics could possibly
distinguish among various scenarios.

\noindent{{\bf Acknowledgement}:
We thank B. Dutta-Roy, H. Gutbroad, 
V. Manko, T. K. Nayak and D. K. Srivastava for useful discussions.} \\


\begin{thebibliography}{99}

\bibitem{pr} J. Alam, S. Raha and B. Sinha, Phys. Rep. {\bf 273}
243 (1996).

\bibitem{arbex} N. Arbex, U. Ornik, M. Plumer, A. Timmermann and R.
Weiner, Phys. Lett. B {\bf 345} 307 (1995).

\bibitem{dks} D. K. Srivastava and B. Sinha, Phys. Rev. Lett. {\bf 73} 
2421 (1994).

\bibitem{dumitru} A. Dumitru, U. Katscher, J. A. Maruhn, H. St\"ocker,
W. Greiner and D. H. Rischke, Phys. Rev. C {\bf 51} 2166 (1995). 

\bibitem{neumann} J. J. Neumann, D. Seibert and G. Fai, Phys. Rev. C
{\bf 51} 1460 (1995).

\bibitem{gale} C. Gale and J. Kapusta Nucl. Phys. B {\bf 357} 65 (1991).

\bibitem{solfrank} J. Solfrank {\it et al}, Phys. Rev. C {\bf 55 } 392 (1997).

\bibitem{karsch} F. Karsch, Z. Phys. C {\bf 38} 147 (1988).

\bibitem{kapusta} J. Kapusta, P. Lichard and D. Seibert, Phys. Rev. D
{\bf 44} 2774 (1991)

\bibitem{halasz} M. A. Halasz, J. V. Steel, G. Q. Li and 
G. E. Brown,  nucl-th/9712006.

\bibitem{xiong} L. Xiong, E. V. Shuryak and G. E. Brown,
Phys. Rev. D {\bf 46} 3798 (1992).

\bibitem{song} C. Song, Phys. Rev. C {\bf 47} 2861 (1993).

\bibitem{ss} S. Sarkar, J. Alam, P. Roy, A. Dutt-Mazumder, B. Dutta-Roy
and B. Sinha, Nucl. Phys. {\bf A634} 206 (1998).

\bibitem{pkr} P. Roy, S. Sarkar, J. Alam and B. Sinha, nucl-th/9803052.

\bibitem{mrho} G. E. Brown and M. Rho, Phys. Rev. Lett. {\bf 66} 2720 (1991).

\bibitem{csong} C. Song and G. Fai, Phys. Rev. C {\bf 58} 1689 (1998).

\bibitem{bjorken} J. D. Bjorken, Phys. Rev. D {\bf 27} 140 (1983).

\bibitem{npa610} P. G. Jones {\it et al}, Nucl. Phys. A {\bf 610} 188c (1996).

\bibitem{wa98con} V. Manko, Int. Nucl. Phys. Conf. (INPC-98), 
August `98,  Paris, France. 

\bibitem{wa80prl} R. Albrecht {et al},  Phys. Rev. Lett. {\bf 76} 3506 (1996).

\end{thebibliography}
\end{document}